\documentclass[letterpaper, 10 pt, conference]{ieeeconf}  

\IEEEoverridecommandlockouts                              
\usepackage{amsmath,amssymb,amsfonts}
\usepackage{hyperref}
\hypersetup{
    colorlinks=true,
    linkcolor=blue,
    filecolor=magenta,      
    urlcolor=cyan,
    pdftitle={Estimation of Dynamic Gaussian Processes},
    pdfpagemode=FullScreen,
    }
\usepackage{amsthm}%
\usepackage{algorithmic}
\usepackage{graphicx}
\usepackage{textcomp}
\usepackage{xcolor}
\def\BibTeX{{\rm B\kern-.05em{\sc i\kern-.025em b}\kern-.08em
    T\kern-.1667em\lower.7ex\hbox{E}\kern-.125emX}}

\usepackage{eso-pic}        

\theoremstyle{definition}
\newtheorem{theorem}{Theorem}

\newtheorem{lemma}[theorem]{Lemma}

\theoremstyle{remark}

\title{\LARGE \bf
Estimation of Dynamic Gaussian Processes}

\author{Jilles van Hulst, Roy van Zuijlen, Duarte Antunes, W.P.M.H. (Maurice) Heemels
\thanks{The research is carried out as part of the ITEA4 20216 ASIMOV project. The ASIMOV activities are supported by the Netherlands Organisation for Applied Scientific Research (TNO) and the Dutch Ministry of Economic Affairs and Climate (project number: AI211006). The research leading to these results is partially funded by the German Federal Ministry of Education and Research (BMBF) within the project ASIMOV-D under grant agreement No. 01IS21022G [DLR], based on a decision of the German Bundestag.}
\thanks{The authors are with the Control Systems Technology Section, Department of Mechanical Engineering, Eindhoven University of Technology, the Netherlands. Emails:{\tt\small \{j.s.v.hulst, r.a.c.zuijlen, d.antunes, m.heemels\}@tue.nl}}%
}

\AddToShipoutPictureBG*{%
	\AtPageUpperLeft{%
		\setlength\unitlength{1in}%
		\hspace*{\dimexpr0.5\paperwidth\relax}
		\makebox(0,-1)[c]{\begin{tabular}{c c}
		        Jilles van Hulst, Estimation of Dynamic Gaussian Processes,\\
				To appear at {\em 62nd IEEE Conference on Decision and Control, CDC 2023}, Singapore, Singapore, 2023.
		\end{tabular}}
}}

\AddToShipoutPictureBG*{%
	\AtPageUpperLeft{%
		\setlength\unitlength{1in}%
		\hspace*{\dimexpr0.5\paperwidth\relax}
		\makebox(0,-21.2)[c]{\begin{tabular}{c c}
				\small \copyright\ 2023 the authors. This work has been accepted to IEEE for publication under a Creative Commons Licence CC-BY-NC-ND.
		\end{tabular}}
}}

\begin{document}

\maketitle
\thispagestyle{empty}
\pagestyle{empty}

\begin{abstract}
Gaussian processes provide a compact representation for modeling and estimating an unknown function, that can be updated as new measurements of the function are obtained. This paper extends this powerful framework to the case where the unknown function dynamically changes over time. Specifically, we assume that the function evolves according to an integro-difference equation and that the measurements are obtained locally in a spatial sense. In this setting, we will provide the expressions for the conditional mean and covariance of the process given the measurements, which results in a generalized estimation framework, for which we coined the term Dynamic Gaussian Process (DGP) estimation. This new framework generalizes both Gaussian process regression and Kalman filtering. For a broad class of kernels, described by a set of basis functions, fast implementations are provided. We illustrate the results on a numerical example, demonstrating that the method can accurately estimate an evolving continuous function, even in the presence of noisy measurements and disturbances.
\end{abstract}

\section{Introduction}
Gaussian Processes (GPs) have seen increasing interest in recent years due to their universal applicability in the estimation of unknown functions. This widely used framework is non-parametric and generally provides accurate estimates with relatively little data. In this paper, we extend this powerful technique to the case when the unknown function \textit{dynamically} changes with \textit{time}.

We consider the estimation of a class of spatio-temporal models, which have a time dimension, as well as a space dimension. The model class includes an evolving function that is described by a generalized version of an integro-difference equation (IDE) \cite{Cressie2011}, \cite{Atluri2019a}. The class also includes an observation equation, which specifies that we take local, noisy measurements of the evolving function at every discrete time step. We call this model class the Dynamic Gaussian Process (DGP).

Our main results pertain to the exact Bayesian estimation of the DGP. We provide analytical expressions for the conditional mean and covariance (with respect to available measurement data and priors). These computations include the calculation of an analytical integral at every time step where new measurements are received. As a result, in the general case, the DGP estimate expression and therefore the computation may become very complex as the number of time steps increases. Therefore, we show that for a restricted class of DGPs, the required computations can be simplified considerably, leading to a rather efficient numerical implementation. The restricted class relies on basis functions and yields computations with complexity $O(M^3)$, with $M$ the number of bases. Importantly, the computation no longer scales with the number of time steps. This is highly beneficial computationally, both when the DGP fits this simplified framework, and when it is approximated to do so. Interestingly, we also show that our Bayesian estimation framework generalizes both Gaussian process regression and Kalman filtering, two of the most important estimation frameworks in the literature.

This is not the first work to examine a connection between the Kalman filter (KF) \cite{Kalman1960} and the Gaussian process (GP) \cite{Rasmussen2006}. In fact, such connections have been studied extensively \cite{Deisenroth2011b}, \cite{Arkka2016}, \cite{Reece2010a}. The computational complexity of a GP estimator can be drastically reduced by incorporating KF logic, given temporal data \cite{Hartikainen2010}, \cite{Todescato2020}. Alternatively, the computation can be simplified by making use of sparse GP methods \cite{Quinonero2005a}, \cite{Liu2020}, recursive methods \cite{Huber2014} or by introducing basis functions \cite{Cressie2008}, \cite{Hensman2018}. GPs have also been used alongside the Kalman filter to estimate state-dependent disturbances \cite{Lee2020}. Moreover, in \cite{Ko2008}, GPs are used to estimate the state dynamics and observation model for different types of the Bayes filter, including the extended Kalman filter.

Turning to the estimation of evolving functions and spatio-temporal models, we find the KF and GP working in tandem. For example, \cite{Sarkka2013} proposes an evolving GP estimator in continuous time using a vector of functions with spatial input, effectively creating an infinite-dimensional KF in continuous time. Moreover, \cite{Wikle1999} developed a framework for the estimation of evolving functions described by IDEs in the case that they are restricted to a set of basis functions. In the current work, however, we also consider the exact Bayesian estimation of the general unrestricted class. Additionally, the current work relaxes specific structure assumptions such as stationary measurement locations which are often not met in practically relevant problems. The Kriged Kalman filter is developed in \cite{Mardia1998}, which presents a Kalman filter projected onto basis functions, making it suitable for the estimation of evolving functions. However, in this work, we establish a rigorous connection to a spatio-temporal model class, which allows our method to estimate IDEs. To the best of the authors' knowledge, there are currently no estimation frameworks that simultaneously generalize the KF and GP-based regression.

The remainder of this paper is organized as follows. In Section \ref{sec:problem}, the main estimation problem of this paper is introduced. Moreover, it is explained how the problem generalizes the estimation problems in GP-based regression and the KF. In Section \ref{sec:method}, the exact solution to the estimation problem is detailed and analyzed. Section \ref{sec:analysis} introduces the special class of DGPs that results in efficient computation. Section \ref{sec:results} presents the results of applying the proposed method in a numerical case study. A code example is available at \url{https://github.com/JvHulst/Dynamic_Gaussian_Processes}. Lastly, Section \ref{sec:conclusions} gives conclusions and recommendations for future research.

\noindent \textbf{Notation.}~Let $\mathbb{R} =(-\infty,\infty)$, and $\mathbb{R}_{\geq0} = [0,\infty)$. Let $\mathbb{N} = \{0,1,\ldots\}$ denote the natural numbers. For $a,b \in \mathbb{N}$, let $\mathbb{N}_{[a,b]} = \{a,a+1,\ldots,b\}$. The identity matrix of size $n$ is denoted by $I_{n}$. Let $\mathbb{S}^n_+:=\{A\in\mathbb{R}^{n\times n} \mid A \succeq 0 \}$ denote the set of symmetric positive semidefinite matrices of size $n \times n$. A normal distribution with mean vector $\bar{m} \in \mathbb{R}^n$ and covariance matrix $\Sigma \in \mathbb{S}^n_+$ is denoted $\mathcal{N}(\bar{m},\Sigma)$. The expected value of a random variable $x \in \mathbb{R}^n$ is denoted $\mathbb{E}\left[x\right]$. The covariance of two random variables is denoted $\text{cov}(x_1,x_2):=\mathbb{E}\left[(x_1-\mathbb{E}\left[x_1\right])(x_2-\mathbb{E}\left[x_2\right])^\top\right]$. A Gaussian process with the mean function $\bar{f}: \mathcal{X} \to \mathbb{R}$ and the covariance function $k: \mathcal{X} \times \mathcal{X} \to \mathbb{R}$ is denoted $\mathcal{GP}(\bar{f}(x),k(x,x'))$.

Given a function $f: \mathcal{X} \to \mathbb{R}$, and a vector $X = [x_1, x_2, \ldots, x_n]^\top \in \mathcal{X}^n$, let $\mathbf{f}(X) = [f(x_1), f(x_2), \ldots, f(x_n)]^\top \in \mathbb{R}^n$. Let $k: \mathcal{X} \times \mathcal{X} \to \mathbb{R}$ be a kernel function, and let $X = [x_1, x_2, \ldots, x_n]^\top \in \mathcal{X}^n$ and $X' = [x'_1, x'_2, \ldots, x'_m]^\top \in \mathcal{X}^{m}$ be vectors. We define the kernel matrix $\mathbf{k}(X,X') \in \mathbb{R}^{n\times m}$ corresponding to $k$ to be the matrix whose ($i$,$j$)-th entry is given by $k(x_i, x'_j)$, i.e.,
$\mathbf{k}(X,X')_{ij} := k(x_i,x'_j),$
for $i \in \mathbb{N}_{[1,n]}$ and for $j \in \mathbb{N}_{[1,m]}$.
\section{Model Class}
\label{sec:problem}
\subsection{Model Class and Problem Statement}
\label{subsec:problem}
The DGP consists of a dynamically evolving function and an observation model. The function at time $t \in \mathbb{N}$ is denoted by $f_t: \mathcal{X} \to \mathbb{R}$, for $x \in \mathcal{X} \subseteq \mathbb{R}$, and its evolution dynamics are described by
\begin{equation}
\label{eq:dynamic_function_system}
	f_{t+1}(x) = \int_\mathcal{X} f_{t}(s) \mu(x,ds) + {w}_{t}(x),
\end{equation}
where $\mu(x,.)$ is a set of measures parametrized by $x$ and such that the integral in \eqref{eq:dynamic_function_system} with respect to the measure $\mu$ is defined as
\begin{equation}
\label{eq:measure_integral}
{\small
\begin{aligned}
    \int_\mathcal{X} f(s) \mu(x,ds) = \int_\mathcal{X} f(s) k_f(x,s) ds + \sum_{i=1}^m f(s_{i}(x))b_i(x){\normalsize.}
\end{aligned}}
\end{equation}
Here, $k_f: \mathcal{X} \times \mathcal{X} \to \mathbb{R}$ is a continuous kernel function, and $s_i: \mathcal{X} \to \mathcal{X}, i \in \mathbb{N}_{[1,m]}$, continuous point mass functions with corresponding continuous weight functions $b_{i}: \mathcal{X} \to \mathbb{R}, i \in \mathbb{N}_{[1,m]}$. We assume that $f_0(x) \sim \mathcal{GP}(\bar{f}_0(x), Q_f(x,x'))$ in which $\bar{f}_0: \mathcal{X} \to \mathbb{R}$ and $Q_f: \mathcal{X} \times \mathcal{X} \to \mathbb{R}$ are, respectively, the mean function and positive semidefinite covariance function of the initial condition function $f_0$, which are both continuous. The disturbances acting on the evolving function are denoted ${w}_t(x) \sim \mathcal{GP}(0,Q_{w}(x,x')),~ t \in \mathbb{N}$, in which $Q_{w}: \mathcal{X} \times \mathcal{X} \to \mathbb{R}$ is a continuous positive semidefinite covariance function. Because the integral transform is a linear transformation, it directly follows that $f_t$ is a GP for any $t \in \mathbb{N}$ \cite{Rasmussen2006}, \cite{Papoulis2002}.

For many of the results presented in this paper, the discrete part of \eqref{eq:measure_integral}, i.e., the sum $\sum_{i=1}^m f(s_i(x)) b_i(x)$, is not considered. This leaves the evolving function dynamics \eqref{eq:dynamic_function_system} as an integro-difference equation (IDE) with a stochastic disturbance. The main purpose of the discrete part is to accompany both GPs and the KF in the same framework. 

While our main results consider first general kernel functions $k_f$, $Q_f$, $Q_{w}$ for Theorems \ref{thm:update} and \ref{thm:prediction} in Section \ref{sec:method} below, they will lead to computationally efficient implementations when the kernels take a special so-called seperable form. We say that a kernel $k(x,x')$ is separable, if
\begin{equation}
    k(x,x')= U^\top(x)\Lambda U(x'),
\end{equation}
for a vector of functions $U(x) := [u_1(x), \ldots, u_M(x)]^\top$ and a matrix $\Lambda \in \mathbb{R}^{M\times M}$. This special form is further examined in Section \ref{sec:analysis}.

The observation model in the DGP is given by
\begin{equation}
\label{eq:function_observation_model}
	Y_t = \mathbf{f}_{t}(X_t) + \mathbf{v}_{t}(X_t),
\end{equation}
where $Y_t \in \mathbb{R}^p$ is a vector of observations of the dynamic function $f_t$ in \eqref{eq:dynamic_function_system} at each time step $t \in \mathbb{N}$, following a vector of corresponding locations $X_t \in \mathcal{X}^p$. $\mathbf{v}_t$ follows from ${v}_t(x) \sim \mathcal{GP}(0, Q_{v}(x,x')),~ t \in \mathbb{N}$, which is noise on the observation, in which the positive semidefinite kernel $Q_{v}: \mathcal{X} \times \mathcal{X} \to \mathbb{R}$ gives the covariance. It is assumed that $f_0(x)$, ${w}_t(x)$, and ${v}_t(x)$ are uncorrelated for any $t, x$.

The problem considered in this paper is the estimation of the function $f_N$ using the data set $\mathcal{D}_N := \{X^{(N)}, Y^{(N)} \}$ in which $X^{(N)} := [X_0^\top, X_1^\top, \ldots, X_N^\top]^\top \in \mathcal{X}^{Np}$ and corresponding $Y^{(N)} := [Y_0^\top, Y_1^\top, \ldots, Y_N^\top]^\top \in \mathbb{R}^{Np}$ as in \eqref{eq:function_observation_model}, and $N \in \mathbb{N}$ is arbitrary. This problem can be motivated from two different perspectives:
\begin{enumerate}
    \item as an extension of the GP-based estimation problem to the case where the Gaussian process evolves in time;
    \item as an extension of the Kalman filter problem to infinite-dimensional systems.
\end{enumerate}
The connection between GP and DGP is shown in Section \ref{subsec:GP} below, while Section \ref{subsec:KF} addresses the connection between the KF and the DGP.


\subsection{Generalization of Gaussian Process Regression}
\label{subsec:GP}
A Gaussian process (GP) is in general an infinite-dimensional set of Gaussian variables, of which any finite subset constitutes a multivariate Gaussian distribution \cite{Rasmussen2006}. The GP framework is often used to estimate unknown functions based on observations that are subject to Gaussian noise, i.e., 
\begin{equation}
\label{eq:GP_function}
	y_t = f(x_t) + {v}_t,
\end{equation}
with $x_t \in \mathcal{X}$ the function's argument, $y_t \in \mathbb{R}$ the observation, $f: \mathcal{X} \to \mathbb{R}$ the unknown function to be estimated, and ${v}_t \sim \mathcal{N}(0,\sigma^2),~ t \in \mathbb{N}$ the Gaussian noise on the observation. An estimate of $f$ from \eqref{eq:GP_function} for $t \in \{0,1,\ldots,N\}$ is completely characterized by the measurements $Y = \{y_0, y_1, \ldots, y_N\} \in \mathbb{R}^N$, the corresponding inputs $X = \{x_0, x_1, \ldots, x_N\} \in \mathcal{X}^N$, and the prior beliefs on its mean and covariance, denoted respectively $\bar{f}: \mathcal{X} \to \mathbb{R}$, and $k: \mathcal{X} \times \mathcal{X} \to \mathbb{R}$. In particular, given a data set $\mathcal{D} := \left\{X,Y\right\}$ of inputs and corresponding outputs according to \eqref{eq:GP_function}, we obtain the following joint distribution
\begin{equation}
\label{eq:GP_distribution}
{\small
	\begin{bmatrix}
	Y\\
	{f}(x)
	\end{bmatrix}
	\sim
	\mathcal{N} \left(
	\begin{bmatrix}
	\mathbf{\bar{f}}(X) \\
	{\bar{f}}(x)
	\end{bmatrix}, \begin{bmatrix}
	\mathbf{k}(X,X) + \sigma^2 I_N & \mathbf{k}(X,x) \\
	\mathbf{k}(x,X) & {k}(x,x)
	\end{bmatrix}
	\right)}.
\end{equation}

We can perform Gaussian process regression by conditioning the joint distribution in \eqref{eq:GP_distribution} with the data $\mathcal{D} = \left\{X,Y\right\}$ using Bayes' rule. The resulting posterior distribution is Gaussian and can be expressed in terms of a mean estimate $\hat{f}: \mathcal{X} \to \mathbb{R}$, and a covariance estimate $\hat{c}: \mathcal{X} \times \mathcal{X} \to \mathbb{R}$:
\begin{equation}
\label{eq:GP_posterior_mean}
	{\hat{f}}(x) := \mathbb{E}[{f}(x) \mid \mathcal{D}] = \bar{f}(x) + \mathbf{L}(x,X)(Y-\mathbf{\bar{f}}(X)),
\end{equation}
\begin{equation}
\label{eq:GP_posterior_covariance}
\begin{aligned}
	{\hat{c}}(x,x') :=&~ \text{cov}\left(f(x),f(x') \mid \mathcal{D}\right) \\
			    =&~ {k}(x,x') - \mathbf{L}(x,X) \mathbf{k}(X,x'),
\end{aligned}
\end{equation}
where $\mathbf{L}(x,X) := \mathbf{k}(x,X) \left[\mathbf{k}(X,X)+\sigma^2 I_N\right]^{-1}$.

The standard GP regression framework can be considered a special case of the problem posed in the previous subsection when \eqref{eq:dynamic_function_system} is restricted to
\begin{equation}
\label{eq:GP_special_case}
    f_{t+1}(x) = f_{t}(x)
\end{equation}
with $f_{0}(x) = f(x)$. The function dynamics \eqref{eq:GP_special_case} are obtained by choosing $k_f=Q_w=0$, $m=1$, $s_{1} = x$, and $b_1 = 1$. In this sense, we can see the stated problem in Section \ref{subsec:problem} as an extension of the GP framework to the case where the unknown function $f$ evolves according to \eqref{eq:dynamic_function_system}. Note that if we choose $Q_w$ arbitrarily instead, the DGP behaves as a GP-based estimator with a forgetting factor.

\subsection{Generalization of the Kalman Filter}
\label{subsec:KF}
The Kalman filter is a well-known algorithm to estimate the state of a dynamical system described by
\begin{equation}
\label{eq:dynamic_state_system}
	x_{t+1} = A x_t + {w}_t,
\end{equation}
with the observation model
\begin{equation}
\label{eq:state_observation_model}
	y_{t} = C_t x_t + {v}_t,
\end{equation}
where $x_t \in \mathbb{R}^{n}$ the state to be estimated, $y_t \in \mathbb{R}^{p}$ the system output, ${w}_t \sim \mathcal{N}(0,{W})$, the disturbance on the state with covariance matrix ${W} \in \mathbb{S}^n_+$, and ${v}_t \sim \mathcal{N}(0,{V})$ the measurement noise with covariance matrix ${V} \in \mathbb{S}^p_+$ at time $t \in \mathbb{N}$. Lastly, $x_0 \sim \mathcal{N}(\bar{x}_0,\bar{S})$ with mean $\bar{x}_0 \in \mathbb{R}^{n}$ and covariance matrix $\bar{S} \in \mathbb{S}^n_+$.

The problem tackled by the Kalman filter is to obtain the estimate of the system's state that minimizes the covariance of the estimation error, given a history of observations $\{y_i\}_{i=0}^N$ generated by \eqref{eq:state_observation_model}. We express the state estimate in terms of a mean $\hat{x}_{t|l} := \mathbb{E}\left[x_t \mid \{y_i\}_{i=0}^l \right]$ and covariance $S_{t|l} := \text{cov}\left(x_t,x_t \mid \{y_i\}_{i=0}^l \right)$. The famous KF result yields equations with which to optimally update these estimates given observations, as well as predict the evolution of the system using the dynamical model \cite{Kalman1960}. The update and prediction steps can be combined into the expressions
\begin{equation}
    \label{eq:Kalman_filter_combined_mean}
	\hat{x}_{t+1|t}=A \hat{x}_{t|t-1} + A L_{t} (y_t-C_t \hat{x}_{t|t-1}),
\end{equation}
and
\begin{equation}
    \label{eq:Kalman_filter_combined_covariance}
	S_{t+1|t} = A S_{t|t-1} A^\top - A L_{t} C_t S_{t|t-1} A^\top + {W},
\end{equation}
with $L_{t} = S_{t|t-1} C_t^\top(C_t S_{t|t-1}C_t^\top+{V})^{-1}$ the Kalman gain. Starting with an initial condition $\hat{x}_{0|-1}=\bar{x}_0$ and $S_{0|-1}=\bar{S}$, we can use the recursive relationships \eqref{eq:Kalman_filter_combined_mean} and \eqref{eq:Kalman_filter_combined_covariance} to obtain an estimate for any $N \in \mathbb{N}$. Importantly, the KF has low memory requirements, due to the compact representation of the estimate. Additionally, the computation scales independently of the number of time steps $N$.

The standard Kalman filter can be seen as the solution to a special case of the estimation problem considered in this paper when we consider the estimation of $f_t$ for a finite set of points $\Xi=\{\xi_1,\xi_2,\ldots,\xi_n\} \in \mathcal{X}^n$. The KF state is then given by $\mathbf{f}_t(\Xi)$. We obtain linear state dynamics by choosing $k_f(x,s) = 0$ $s_{i} = \xi_i$, and $b_i(\xi_j) = A_{ij}$ in \eqref{eq:dynamic_function_system}, with $A_{ij}$ the elements of $A$ in \eqref{eq:dynamic_state_system} for $i,j \in \mathbb{N}_{[1,n]}$. Lastly, the observation model \eqref{eq:state_observation_model} is a result of the sampling locations $X_t$ in \eqref{eq:function_observation_model}.

The following section details the solution to the DGP estimation problem.

\section{Estimation of Dynamic Gaussian Processes}
\label{sec:method}
Here, we consider the estimation problem for the system \eqref{eq:dynamic_function_system} with observation model \eqref{eq:function_observation_model}. To solve the DGP estimation problem, we will first state two theorems, which together constitute the main contribution of this paper. The theorems conceptually generalize Theorem 4.1 in \cite[Chapter 7]{Astrom1970}, related to the Kalman filter.

We aim to estimate the true evolving function $f_{N}(x)$ for arbitrary $N \in \mathbb{N}$ using the data set $\mathcal{D}_N$, and the model \eqref{eq:dynamic_function_system} with the continuous part only, i.e., 
\begin{equation}
\label{eq:dynamic_function_system_IDE}
	f_{t+1}(x) = \int_\mathcal{X} f_{t}(s) k_f(x,s) ds + {w}_{t}(x).
\end{equation}
The estimate is characterized by a mean function
\begin{equation}
\label{eq:mean_function_definition}
    \hat{f}_{t|l}(x) := \mathbb{E}\left[f_t(x) \mid \mathcal{D}_l\right],
\end{equation}
and a covariance function
\begin{equation}
\label{eq:covariance_function_definition}
    \hat{c}_{t|l}(x,x') := \text{cov}\left(f_t(x),f_t(x') \mid \mathcal{D}_l\right).
\end{equation}

The first theorem is concerned with the conditioning of the mean and variance estimate based on newly obtained data.
\begin{theorem}[\textbf{DGP update}]
\label{thm:update}
    Consider the evolving function dynamics \eqref{eq:dynamic_function_system_IDE} with a prior belief of $f_{t}$ based on $\mathcal{D}_{t-1}$, expressed in terms of a mean $\hat{f}_{t|t-1}$ and a covariance $\hat{c}_{t|t-1}$ as defined in \eqref{eq:mean_function_definition} and \eqref{eq:covariance_function_definition}. Consider the updating of this belief based on new data $X_t$, $Y_t$, obtained according to \eqref{eq:function_observation_model}. The posterior mean and covariance, conditioned on $X_t$ and $Y_t$, are then given by
    \begin{equation}
    \label{eq:DGP_mean_update}
        \hat{f}_{t|t}(x) = \hat{f}_{t|t-1}(x) + \mathbf{L}_t(x,X_t) (Y_t - \mathbf{\hat{f}}_{t|t-1}(X_t)),
    \end{equation}
    \begin{equation}
    \label{eq:DGP_covariance_update}
        \hat{c}_{t|t}(x,x') = \hat{c}_{t|t-1}(x,x')-\mathbf{L}_t(x,X_t)\mathbf{\hat{c}}_{t|t-1}(X_t,x'),
    \end{equation}
    in which 
    {\small
    \begin{equation*}
        \mathbf{L}_t(x,X_t) = \mathbf{\hat{c}}_{t|t-1}(x,X_t) \left[\mathbf{\hat{c}}_{t|t-1}(X_t,X_t) + \mathbf{Q_{v}}(X_t,X_t) \right]^{-1}.
    \end{equation*}}\noindent
\end{theorem}
\begin{proof}
The first step in the proof pertains to the separation of the new data $X_t$, $Y_t$ from the previously obtained data $\mathcal{D}_{t-1}$ in terms of their contribution to the belief. To do so, we define $\tilde{Y}_t := Y_t - \mathbb{E}\left[Y_t \mid \mathcal{D}_{t-1}\right],$ which is independent from $\mathcal{D}_{t-1}$, following Theorem 3.2 in \cite[Chapter 7]{Astrom1970}. Note that since we only consider finite-dimensional data here, the result from \cite{Astrom1970} can directly be applied, even though we are considering an infinite-dimensional system. We obtain
{\small
\begin{equation*}
    \begin{aligned}
        \mathbb{E}\Bigl[f_t(x) &\bigm\vert \mathcal{D}_{t-1}, X_t, Y_t \Bigr] = \mathbb{E}\left[f_t(x) \bigm\vert \mathcal{D}_{t-1}, X_t, \tilde{Y}_t\right] \\
        =&~ \hat{f}_{t|t-1}(x) + \mathbb{E}\left[ f_t(x) \bigm\vert X_t, \tilde{Y}_t\right] - \mathbb{E}\left[f_t(x)\right].
    \end{aligned}
\end{equation*}}\noindent
The first equality holds as it is a transformation of variables. The second holds due to independence of $X_t$ and $\tilde{Y}_t$ with $\mathcal{D}_{t-1}$, following Theorem 3.3 in \cite[Chapter 7]{Astrom1970}. The expression shows that the incorporation of newly obtained data $X_t$, $Y_t$ can be performed as an update on $\hat{f}_{t|t-1}$.

Next, we can express the second term in the last expression by conditioning $f_t$ on the data $X_t$ and $\tilde{Y}_t$ using Bayes' rule. We have
{\small
\begin{equation*}
    \mathbb{E}\left[ f_t(x) \bigm\vert X_t, \tilde{Y}_t\right] = \mathbb{E}\left[f_t(x)\right] + \mathbf{L}_t(x,X_t)(\tilde{Y}_t),
\end{equation*}}\noindent
in which 
{\small
\begin{equation*}
    \mathbf{L}_t(x,X_t) = \mathbf{\hat{c}}_{t|t-1}(x,X_t) \left[\mathbf{\hat{c}}_{t|t-1}(X_t,X_t) + \mathbf{Q_{v}}(X_t,X_t) \right]^{-1}.
\end{equation*}}\noindent
To get the posterior covariance, we use the definition of the covariance and simply multiply the estimation error $e_{t|t} := f_t-\hat{f}_{t|t}$ with its transpose to obtain
{\small
\begin{equation*}
    \hat{c}_{t|t}(x,x') = \hat{c}_{t|t-1}(x,x')-\mathbf{L}_t(x,X_t)\mathbf{\hat{c}}_{t|t-1}(X_t,x'),
\end{equation*}}\noindent
which completes the proof.
\end{proof}

We now proceed with the second theorem, which details the propagation of the mean and covariance estimates through the function dynamics \eqref{eq:dynamic_function_system_IDE}.
\begin{theorem}[\textbf{DGP prediction}]
\label{thm:prediction}
    Consider the evolving function dynamics given by \eqref{eq:dynamic_function_system_IDE} with a prior belief about the mean and variance of $f_t$ based on the information set $\mathcal{D}_t$, denoted $\hat{f}_{t|t}$, and $\hat{c}_{t|t}$, respectively, as in \eqref{eq:mean_function_definition} and \eqref{eq:covariance_function_definition}. Then, the belief evolves due to the functional dynamics in \eqref{eq:dynamic_function_system_IDE} to obtain
    \begin{equation}
    \label{eq:DGP_mean_prediction}
        \hat{f}_{t+1|t}(x) = \int_\mathcal{X} k_f(x,s)\hat{f}_{t|t}(s) ds,
    \end{equation}
    and
    \begin{equation}
    \label{eq:DGP_covariance_prediction}
    {\footnotesize
    \begin{aligned}
        \hat{c}_{t+1|t}(x,x') = \int_\mathcal{X} \int_\mathcal{X} k_f(x,s)\hat{c}_{t|t}(s,s')k_f(s',x') ds ds' + Q_{w}(x,x'){\normalsize.}
    \end{aligned}}
    \end{equation}\noindent
\end{theorem}
\begin{proof}
Starting with the mean estimate, we have
{\small
\begin{equation*}
\begin{aligned}
    \mathbb{E}\left[f_{t+1}(x) \mid \mathcal{D}_t\right] =&~ \mathbb{E}\left[\int_\mathcal{X} k_f(x,s)f_t(s) ds + w_t(x) \Bigm\vert \mathcal{D}_t\right], \\
    =&~ \int_\mathcal{X} k_f(x,s)\mathbb{E}\left[f_t(s) \mid \mathcal{D}_t\right] ds, \\
\end{aligned}
\end{equation*}}\noindent
where the first equality follows from the dynamics \eqref{eq:dynamic_function_system_IDE}, and the second follows from the zero-mean assumption on $w_t$ and from the linearity of the expectation, since the integral transform is a linear map.

Next, the covariance estimate. We first define the deviation from the expected mean as $e_{t+1|t} := f_{t+1}-\hat{f}_{t+1|t}$.
We have 
{\small
\begin{equation*}
\begin{aligned}
    \text{cov} ( f_{t+1}&(x),f_{t+1}(x') \mid \mathcal{D}_t ) = \mathbb{E} \left[ e_{t+1|t}(x)e_{t+1|t}(x') \mid \mathcal{D}_t \right]\\
    =&~ \mathbb{E}\Biggl[w_t(x)w_t(x') + \left(\int_\mathcal{X} k_f(x,s)\left(f_{t}(s)-\hat{f}_{t|t}(s)\right) ds\right) \\
    & \quad \times \left(\int_\mathcal{X} k_f(x',s')\left(f_{t}(s')-\hat{f}_{t|t}(s')\right) ds'\right ) \Bigm\vert \mathcal{D}_t\Biggr] \\
    =&~ \int_\mathcal{X} \int_\mathcal{X} k_f(x,s)\hat{c}_{t|t}(s,s')k_f(s',x') ds ds' + Q_{w}(x,x').
\end{aligned}
\end{equation*}}\noindent
The first equality holds by definition of the covariance. The second holds because $w_t$ is independent of both $f_{t}$ and $\hat{f}_{t|t}$. The last equality holds because the integral transform is linear, because of the independence between $w_t$ and $\mathcal{D}_t$, and by definition of $\hat{c}_{t|t}$ and $Q_{w}$. This completes the proof.
\end{proof}

We can combine the theorems to estimate the DGP at arbitrary time steps $N \in \mathbb{N}$. To do so, we initialize the estimator's mean and covariance, ideally, but not necessarily, as $\hat{f}_{0|-1} = \bar{f}_0$, and $\hat{c}_{0|-1} = Q_f$. Then we perform the following algorithm at each time step $t$:
\begin{enumerate}
    \item Update the DGP estimator using Theorem \ref{thm:update} according to newly obtained data;
    \item Predict the DGP evolution using Theorem \ref{thm:prediction}.
\end{enumerate}
These steps can be performed recursively until we obtain $\hat{f}_{N|N}$ and $\hat{c}_{N|N}$, which presents the solution to the problem posed in Section \ref{sec:problem}.

The implementation of Theorems \ref{thm:update} and \ref{thm:prediction} applies to the estimation of DGPs with a general structure. However, it may get computationally very expensive to perform as $N$ increases due to the continuous integral, which in general produces complicated expressions for the mean and covariance estimates.

The next section details how the computation of the estimator is reduced when the kernels of the DGP are separable.

\section{Separable Kernels}
\label{sec:analysis}
In this section, we show that when $k_f$, $Q_f$, and $Q_{w}$ are separable kernels we can perform the required computations of the previous section efficiently. We start with exact DGP when the kernels are truly separable, and detail afterward that separable kernels can be used to approximate the problem from Section \ref{subsec:problem}. To show the reductions in computational steps in case of separability of kernels $k_f$, $Q_f$, and $Q_{w}$, consider the following lemmas.
\begin{lemma}
\label{lem:seperable_kernels}
Suppose that
\begin{equation}
\label{eq:seperable_assumption}
    \begin{aligned}
        k_f(x,x') =&~ U^\top(x) \Lambda U(x'),\\
        Q_f(x,x') =&~ U^\top(x) \Lambda_f U(x'),\\
        Q_{w}(x,x') =&~ U^\top(x) \Lambda_w U(x')
    \end{aligned}
\end{equation}
with $\Lambda \in \mathbb{R}^{M \times M}$, $\Lambda_f, \Lambda_w \in \mathbb{S}_+^M$ and $U(x) := [u_1(x), u_2(x), \ldots, u_M(x)]^\top$ a vector of basis functions in which $u_i: \mathcal{X} \to \mathbb{R}, i \in \{1,2,\ldots,M\}$ continuous, then the DGP covariance update step in \eqref{eq:DGP_covariance_update} reduces to
\begin{equation}
\label{eq:reduced_DGP_covariance_update}
    {\hat{c}}_{t|t}(x,x') =  {U}^\top(x) \Psi_{t|t} {U}(x')
\end{equation}
with $\Psi_{t|t} := \Psi_{t|t-1} + \Gamma_t \mathbf{U}^\top(X_t)\Psi_{t|t-1},$ in which $\Psi_{0|-1}=\Lambda_f$, and
{\small
\begin{equation*}
    \Gamma_t := \Psi_{t|t-1}\mathbf{U}(X_t) \bigl[\mathbf{U}^\top(X_t)\Psi_{t|t-1}\mathbf{U}(X_t) + \mathbf{Q_{v}}(X_t,X_t)\bigr]^{-1}
\end{equation*}}\noindent
with $\mathbf{U}(X_t) := \begin{bmatrix}
    \mathbf{u}_1(X_t) &
    \mathbf{u}_2(X_t) &
    \ldots            &
    \mathbf{u}_1(X_t)
\end{bmatrix}^\top \in \mathbb{R}^{M \times p}.$
Additionally, the covariance prediction step in \eqref{eq:DGP_covariance_prediction} is reduced to
\begin{equation}
\label{eq:reduced_DGP_covariance_prediction}
    {\hat{c}}_{t+1|t}(x,x') = {U}^\top(x)\Psi_{t+1|t}{U}(x')
\end{equation}
with $\Psi_{t+1|t} := \Lambda \Lambda_U \Psi_{t|t} \Lambda_U \Lambda^\top + \Lambda_w$, in which $\Lambda_U := \int_\mathcal{X} U(x)U^\top(x) dx$.
\end{lemma}
\begin{proof}
Initialize $\hat{c}_{0|-1}(x,x') =  U^\top(x) \Psi_{0|-1} U(x')$ with $\Psi_{0|-1} = \Lambda_f$, then substitute \eqref{eq:seperable_assumption} into \eqref{eq:DGP_covariance_update} and \eqref{eq:DGP_covariance_prediction}.
\end{proof}
\noindent \textit{Remark.} Note that $Q_{v}$ does not need to be separable, as the measurement noise function ${v}_t$ is always evaluated for a finite set of spatial locations $X_t$ when taking observations according to \eqref{eq:function_observation_model}, and $\mathbf{Q_v}(X_t,X_t)$ can be simply computed and used in the calculations above.
    
To state \textit{Lemma} \ref{lem:seperable_kernels} in other words, the assumption in \eqref{eq:seperable_assumption} causes the covariance estimate $\hat{c}_{t|l}$ to remain in the basis under operation of both the update and the prediction steps. This enables a computationally efficient implementation of the estimator since we now only need to keep track of the matrix $\Psi_{t|l}$. However, the mean estimate computation still scales poorly as the time index $N$ increases. Efficient implementation of the mean estimate is enabled by the application of the following lemma.
\begin{lemma}
\label{lem:prior_mean}
    If \eqref{eq:seperable_assumption} holds and additionally
    \begin{equation}
        \label{eq:DGP_prior_mean_assumption}
        \bar{f}_0(x) = U^\top(x)\bar{z},
    \end{equation}
    with $\bar{z} \in \mathbb{R}^M$ and $U$ the same vector of functions as in \eqref{eq:seperable_assumption}, then the update of the mean estimate \eqref{eq:DGP_mean_update} reduces to
    \begin{equation}
    \label{eq:reduced_DGP_mean_update}
        {\hat{f}}_{t|t}(x) = {U}^\top(x) z_{t|t},
    \end{equation}
    with $z_{t|t} := z_{t|t-1} + \Gamma_t(Y_t-\mathbf{U}^\top(X_t)z_{t|t-1}).$ Additionally, the prediction step in \eqref{eq:DGP_mean_prediction} is reduced to
    \begin{equation}
    \label{eq:reduced_DGP_mean_prediction}
        {\hat{f}}_{t+1|t}(x) = {U}^\top(x) z_{t+1|t},
    \end{equation}
    with $z_{t+1|t} = \Lambda \Lambda_U z_{t|t}.$
\end{lemma}
\begin{proof}
Initialize $\hat{f}_{0|-1}(x) =  U^\top(x) z_{0|-1}$ with $z_{0|-1} = \bar{z}$, then substitute \eqref{eq:seperable_assumption} into \eqref{eq:DGP_mean_update} and \eqref{eq:DGP_mean_prediction}.
\end{proof}

Note the similarity of the expressions in the lemmas to the standard KF. With \textit{Lemma \ref{lem:seperable_kernels}} and \textit{Lemma \ref{lem:prior_mean}}, the DGP can be interpreted as a Kalman filter where the state estimates are projected onto the functions in $U$, similar to \cite{Mardia1998}. Additionally, it can be evaluated at every spatial location and the measurement data can occur anywhere in the set $\mathcal{X}$.

If both lemmas apply, the problem presented in Section \ref{sec:problem} can be exactly estimated using the efficient equations presented in this section. For the more general case, where \eqref{eq:dynamic_function_system} does not satisfy the lemmas, the problem can be approximated using basis functions. In order to do this, we approximate the function $\bar{f}_0$ in the basis vector $U$, and the kernel functions $k_f$, $Q_f$, $Q_{w}$ in the basis vector $U(x) \otimes U(x')$, with $\otimes$ the Kronecker product. We can find $\Lambda$, $\Lambda_f$, $\Lambda_w$ and $\bar{z}$ by performing standard parametric function approximation methods such as least squares using the $L^2$-norm of the approximation error. Note that this least squares problem can be numerically approximated using Riemann sums. Two important choices for the basis structure in $U$ are highlighted. The first is discrete bins, while the second is the Fourier basis. Note that the class of separable kernels is dense in the space of ${L}^2$ integrable kernels for both of these choices. 

The key benefit of the lemmas lies in the reduced computational complexity. When the DGP structure admits the basis functions, either through approximation or exactly, the computation scales with $O(M^3)$, similar to the standard KF. Additionally, this implies that the approximation framework provides a choice to the user in regard to computational burden.

\section{Results}
\label{sec:results}
\begin{figure}[!t]
    \centering
    \includegraphics[width=80mm]{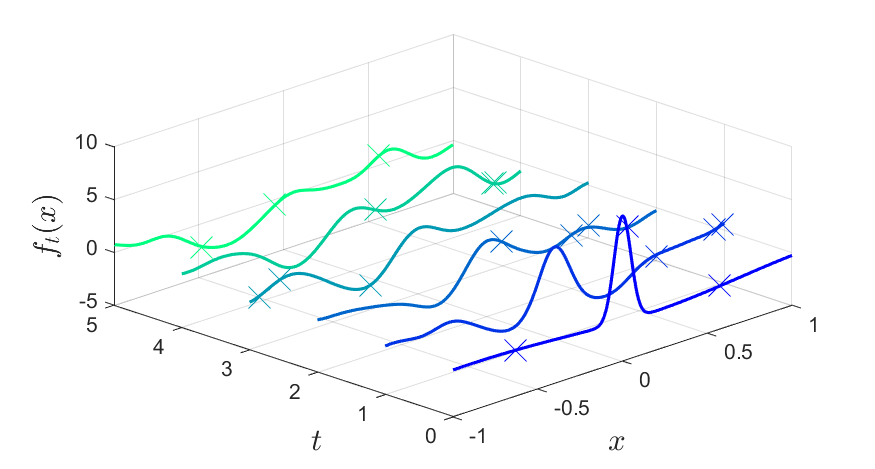}
    \caption{Function $f_t$ over time steps $t$ using a smoothening evolution kernel $k_f$, a Gaussian initial mean function $\bar{f}$, a smooth initial condition covariance $Q_f$, and smooth zero-mean disturbances $w_t$.}
    \label{fig:true_func}
\end{figure}

\begin{figure}[!t]
    \centering
    \includegraphics[width=80mm]{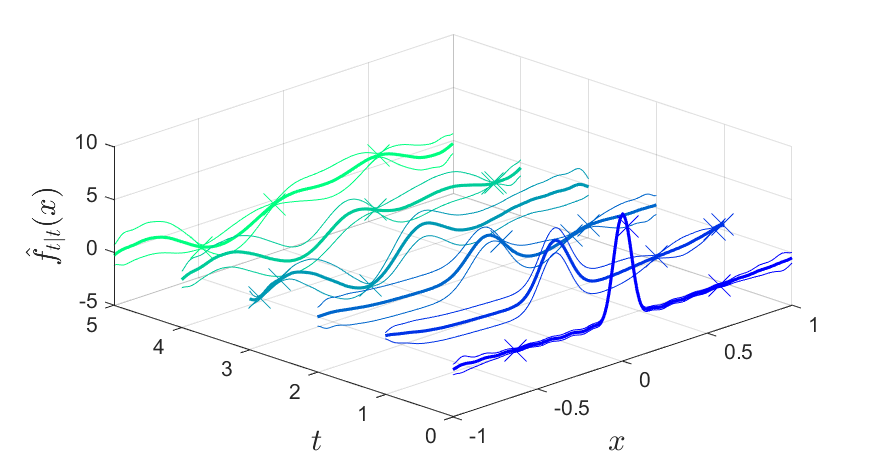}
    \caption{Estimate $\hat{f}_{t|t}$ of function $f_t$ using 31 Fourier bases over time steps $t$. A confidence interval of 95\% is plotted alongside the mean estimate.}
    \label{fig:approx_func}
\end{figure}

\begin{figure}[!t]
    \centering
    \includegraphics[width=80mm]{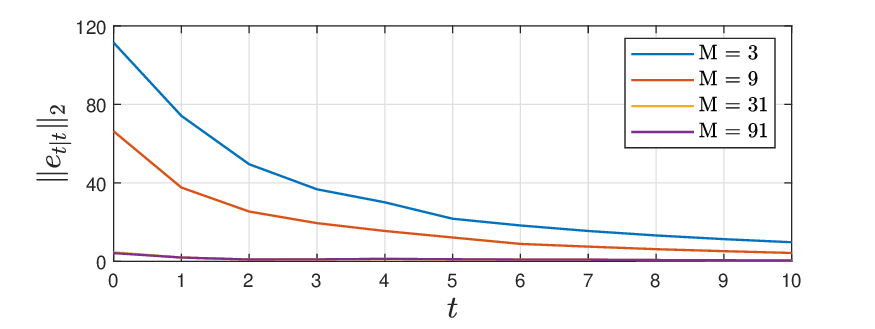}
    \caption{2-norm of the estimation error $e_{t|t}$ over time steps $t$ without disturbances. A comparison is made between 3, 9, 31, and 91 Fourier bases in $U$.}
    \label{fig:error_2_norm}
\end{figure}

In this section, an application of DGPs and of the proposed approximation method presented in Section \ref{sec:analysis} to a numerical example is presented. To assess the estimation accuracy of the approximate methods, a ground truth model is considered, which uses 625 basis functions in the form of discrete bins compared to up to 91 Fourier bases in the approximation. We select $\mathcal{X} = [-1,1]$. We model
{\small
\begin{equation*}
    k_f(x,x') = \phi(5.13,0.07,(x-x')),
\end{equation*}}\noindent
in which $\phi(a,\sigma,d) := a e^{-\frac{\lvert d\rvert^2}{2 \sigma^2}}$ is a parametrized squared exponential kernel. This choice results in local smoothing and decay of the function $f_t$ every time step $t$. 
Furthermore,
{\small
\begin{equation*}
    Q_f(x,x') = \phi(1,0.7,(x-x')),
\end{equation*}}\noindent
which ensures that the initial condition $f_0$ smoothly deviates from its mean. The disturbances are modeled to have smooth covariance using
{\small
\begin{equation*}
    Q_{w}(x,x') = \phi(0.35,0.15,(x-x')).
\end{equation*}}\noindent
Next,
{\small
\begin{equation*}
    Q_{v}(x,x') = \sigma_v^2 \delta(x-x') 
\end{equation*}}\noindent
with $\delta$ the Dirac delta function and $\sigma_v = 0.1$ simulates white noise on the measurements since this choice implies there is no spatial correlation in ${v}_t$. Lastly,
{\small
\begin{equation*}
    \bar{f}_0(x) = \phi(10,0.05,x)
\end{equation*}}\noindent
simulates a Gaussian in the center of the function domain at the initial time step $t=0$. The spatial locations $X_t$ are selected randomly every $t$ using a uniform distribution over $\mathcal{X}$ with $p=3$. Note that now, $X_t$ is a random variable. This still allows us to use the theorems and lemmas, however, as it is independent of $\mathcal{D}_{t-1}$. The results are shown in Fig.~\ref{fig:true_func}, Fig.~\ref{fig:approx_func}, and Fig.~\ref{fig:error_2_norm}.

Observe Fig.~\ref{fig:true_func}, which shows a simulation of the ground truth model. Note the smoothing of the function as the time index $t$ increases as a result of the choice of kernel $k_f$. Note furthermore that disturbances affect the function during the entire simulation, for instance at the location $x=-0.3$ at time step $t=1$.

Fig.~\ref{fig:approx_func} displays the approximation of the DGP in Fig.~\ref{fig:true_func} using 31 Fourier bases. Note that wherever an observation occurs, indicated by a cross, the local confidence is improved. Note furthermore that even in the absence of measurements, the estimate changes due to the prediction of the dynamics. Finally, observe that the local disturbance in the true function at $t=1$ only appears in the estimate at $t=3$, because the local observation $X_t$ did not occur in that local area immediately when the disturbance happened.

Lastly, consider Fig.~\ref{fig:error_2_norm}. More bases yield a lower error 2-norm at $t=0$, since the initial mean function is approximated better as the number of bases increases, with diminishing returns. Note that even when the number of bases is high, the error at $t=0$ is nonzero due to the uncertainty function's initial condition. After some measurements, this uncertainty is resolved and the error decreases. As the function evolves, it smoothens out, which reduces the amount of high-frequency content, making it easier to approximate even when the number of bases is lower.

\addtolength{\textheight}{-3cm}   
\section{Conclusions}
\label{sec:conclusions}
This paper presented an exact method for the Bayesian estimation of dynamic Gaussian processes (DGPs) describing dynamically evolving uncertain functions. An explicit connection to the powerful techniques of Kalman filtering and Gaussian process regression is made. It is also shown that the computation of the DGP estimation is reduced when a separability structure is present in the kernel and mean functions describing the problem. Next to this exact method, we exploited the separable structure to derive an approximation method using basis functions with reduced memory and computational burden during operation. The proposed framework is demonstrated in a numerical case study.

Future research avenues include the analysis of the error resulting from the approximation, the extension of the problem to include control inputs, and connecting the framework to the estimation of partial differential equations.


\bibliographystyle{IEEEtran}
{\small
\bibliography{CDC2023}
}

\end{document}